\documentclass[floatfix,preprint]{revtex4}
\usepackage{graphicx}
\usepackage{amsmath}
\usepackage{latexsym}
\usepackage{color}
\usepackage{epsfig}
\usepackage{setspace}


\begin{document}

\title{VUV frequency combs from below-threshold harmonics}

\author{Dylan C. Yost$^1$, Thomas R. Schibli$^1$, Jun Ye$^{1,*}$, Jennifer L. Tate$^2$, James Hostetter$^2$, Mette B. Gaarde$^2$ and Kenneth J. Schafer$^2$}
\affiliation{$^1$JILA, National Institute of Standards and Technology, and University of Colorado. \\ Department of Physics, University of Colorado, Boulder,
Colorado 80309-0440
\\$^2$Department of Physics and Astronomy, Louisiana State University,
Baton Rouge, Louisiana 70803-4001
}

\maketitle
\textbf{Recent demonstrations of high-harmonic generation (HHG) at very high repetition frequencies ($\sim$100 MHz)~\cite{Jones1, Gohle} may allow for the revolutionary transfer of frequency combs~\cite{Cundiff, Udem, Marian} to the vacuum ultraviolet (VUV).  This advance necessitates unifying optical frequency comb technology with strong-field atomic physics~\cite{Lewenstein94, Meckel, Hentschel, Sansone}. While strong-field studies of HHG have often focused on above-threshold harmonic generation (photon energy above the ionization potential), for VUV frequency combs an understanding of below-threshold harmonic orders and their generation process is crucial.  Here we present a new and quantitative study of the harmonics 7-13 generated below and near the ionization threshold in xenon gas. We show multiple generation pathways for these harmonics that are manifested as on-axis interference in the harmonic yield. This discovery provides a new understanding of the strong-field, below-threshold dynamics under the influence of an atomic potential and allows us to quantitatively assess the achievable coherence of a VUV frequency comb generated through below threshold harmonics. We find that under reasonable experimental conditions temporal coherence is maintained. As evidence we present the first explicit VUV frequency comb structure beyond the 3rd harmonic~\cite{Jones1, Gohle, Witte2}.}

High repetition frequency HHG has recently been enabled through the use of femtosecond enhancement cavities~\cite{Jones1, Gohle}.  In addition to potentially transferring frequency comb techniques to the VUV, the enhancement cavity technique allows for greater harmonic photon flux and a near-perfect Gaussian fundamental beam which provides for exceptionally clean and high signal-to-noise studies of the HHG process itself.  In this Report, we utilize these techniques and present a new set of experimental and theoretical studies of harmonics 7 through 13, generated below and near the ionization threshold in xenon gas with an intense 1070 nm laser pulse.  These harmonic orders are of great interest for the development of VUV frequency combs and it is crucial to understand how fluctuations in the driving laser intensity influence the pulse-to-pulse coherence properties of the train of VUV pulses through the non-linear HHG process.

Surprisingly, we find that harmonics as low as the 7th contain multiple contributions with different intensity-dependent phases. Our laser system permits a clean observation of interference between these phase contributions at unprecedented signal-to-noise ratios.
In particular, we see experimentally a strong contribution from a component that has a large, intensity-dependent phase and we show theoretically that it originates in semi-classical laser-driven continuum dynamics. We demonstrate that this phase is significantly larger than for above threshold harmonics due to the influence of the atomic potential on these low order harmonics. These measurements also provide a quantitative basis to evaluate the long-term temporal coherence of the VUV pulse train, a necessary condition for the establishment of a VUV frequency comb. Under reasonable experimental conditions, we find that the frequency comb structure should indeed emerge.  To verify this, we demonstrate for the first time coherence between successive pulses in our 7th harmonic pulse train ($\lambda \approx$ 153 nm), which shows a coherence time $\sim$5 orders of magnitude larger than that shown previously~\cite{Cavalieri, Lynga}.

For this study, we utilize a passive optical cavity to enhance a high-power mode-locked femtosecond fiber laser at 1070 nm. We are able to achieve the peak power necessary for HHG at high repetition frequencies (136 MHz), which gives a very high signal to noise ratio and enables the pulse-to-pulse coherence measurements shown later.  Figure 1 shows the experimentally measured on-axis yields of harmonics 7 through 13 as a function of the laser intensity in the center of the gas jet.  Interestingly, all four harmonics exhibit complicated intensity-dependent yields, with steep increases interrupted by steps.  For harmonics 7 and 11 the intensity steps are more pronounced and occur at lower intensities than for harmonics 9 and 13. The insets show experimentally measured far-field spatial profiles of each harmonic at an intensity of $~2 \times 10^{13}$ W/cm$^2$. At this intensity, harmonics 7 and 11 also exhibit strong off-axis halos in their spatial distributions whereas 9 and 13 do not. We also note that, in contrast to on-axis, the off-axis intensity-dependent yield displays no interference effects.

The experimental results are very well reproduced by the theoretical calculations (shown in Fig.~1 for comparison) obtained via  the coupled, non-adiabatic solutions of the time-dependent Schr\"odinger equation (TDSE) and the wave equation for a gas of xenon atoms exposed to an intense, 1070~nm laser pulse. Our approach is described in detail in~\cite{GaardeReview}, with the important difference that for the work described here we are directly integrating the TDSE numerically within the single active electron approximation~\cite{Schafer97}. We are thus treating the laser electric field and the atomic potential on an equal footing which is necessary to describe harmonics with photon energies below and close to the ionization threshold. As initial conditions for the calculation we use the same parameters as the experiment~\cite{footnote2}. The theory results reproduce the overall increase of the yield with intensity and exhibit prominent intensity-dependent steps in harmonics 7 and 11, and less pronounced steps in harmonics 9 and 13. In addition, the positions of the steps are remarkably well reproduced by theory.

Our experimental and theoretical results suggest that the observed intensity dependence  is due to interference between contributions to the harmonic generation process which have different intensity-dependent phases. To explore these different contributions, we analyze the single atom intensity-dependent dipole moment $d_q(I)$ for each harmonic $q$ in terms of its conjugate phase variable $\alpha$, as described in detail in~\cite{BalcouQpath}. From the calculated dipole phase we deduce the weight of each contribution, called a quantum path in the literature, over a range of intensities around $I_0$, via the transform:
\begin{equation}
{\tilde d}_q(\alpha) = \int d_q(I) e^{i\alpha U_p(I)/\hbar \omega}W(I-I_0) dI,
\end{equation}
where $W(I-I_0)$ is a window function centered on $I_0$~\cite{BalcouQpath}. The contribution characterized by phase-coefficient $\alpha_j$ has a phase which is proportional to the intensity through $\phi_j = \alpha_j U_p(I)/\hbar\omega$, where $U_p = I/4\omega^2$ is the ponderomotive energy in atomic units, and $\omega$ is the laser frequency. For above-threshold harmonics, the phase coefficients obtained in this way correspond well to those predicted by the semi-classical model~\cite{GaardeQpath}, including the familiar short and long trajectories with phase coefficients of $\alpha_1\approx 0.2\pi$ and $\alpha_2\approx 2\pi$ respectively~\cite{Lewenstein95}. To our knowledge, this analysis has not been applied to below-threshold harmonics before.

Figure 2 shows the results of the quantum path analysis for harmonics $7\!-\!13$ in xenon. All the harmonics exhibit multiple quantum path contributions; the two dominant ones have phase coefficients $\alpha_0 \approx 0$ and $\alpha_2 \approx 2.5\pi\!-\!3\pi$.
To understand the origin of the $\alpha_2$ contribution to the below-threshold harmonics, we have studied electron trajectories in a generalized semi-classical model in which the atomic potential is present. Crucially, we have found that in the presence of the atomic potential, low-energy electrons can lose enough energy to have less than zero total energy at $x=0$ (the position of the ion), leading to the emission of below-threshold harmonics. In the tunnel ionization model, this can only happen for electrons that have followed the long trajectory because they return to $x_i$, the position of ionization, at a time when the force from the laser field is against their motion. We find that the long trajectories accumulate an intensity-dependent phase that is larger than in the no-potential case, with a phase coefficient close to $3\pi$ for the range of intensities we are interested in, rather than the $ 2\pi$ predicted with no potential. This is in good agreement with the result in Fig. 2. There are no  negative-energy returns for short-trajectory electrons in the tunneling model because they return to $x_i$ at a time when the laser field further accelerates them toward $x=0$.  This leads us to designate the $\alpha_0$ contribution as a multiphoton process with no intensity-dependent phase, and not a generalized short trajectory. We note that harmonics 3 and 5 show only an $\alpha_0$ contribution.

The calculations in Fig. 2 confirm our interpretation of the experimental results in Fig. 1 as manifestations of the multiple generation mechanisms. The strong halos in the far-field spatial profiles of harmonics 7 and 11 are consistent with a large $\alpha_2$ contribution to these harmonics at intensities below $2\times 10^{13}$~W/cm$^2$. This is because the large intensity-dependent phase, coupled with the Gaussian spatial profile of the fundamental beam, produces a strongly curved harmonic wavefront~\cite{Bellini}. Although multiple generation pathways are familiar from HHG high above the ionization threshold, it has never been observed (or predicted) so far below threshold. Previous works have noted that harmonics close to threshold can have large intensity dependent phases~\cite{Peatross}, and that harmonics well below threshold can exhibit a step in the intensity-dependent yield, which was interpreted as an atomic resonance effect~\cite{Balcou}. Our results explain and extend these observations to harmonics well below and around threshold without invoking atomic resonances. Since it turns out that the intensity dependence of the dipole phase below threshold is due primarily to continuum dynamics, we have now a general quantitative connection between the laser intensity and the phase for these harmonics.

This discovery allows us to connect these measurements to several other strong field phenomena observed in recent years. These experiments, which measured ionization dynamics rather than photon emission, all invoke motion in the combined field of the ion and laser field to explain the behavior of electrons with energies below the ionization threshold~\cite{Ho2003, Nubbemeyer2008, Shuman2008}.  Our observations, and the high dynamic range associated with them, open the possibility to study the electron dynamics invoked in these experiments directly via the HHG process.

Since the HHG process is extremely nonlinear, it has been a significant concern that frequency comb coherence could not be maintained through high harmonic generation.  Previous experiments have measured the coherence maintained in HHG but only over the time scale of a single pulse~\cite{Cavalieri, Lynga}.  However, to produce a frequency comb, it is a prerequisite that coherence is maintained between successive pulses in the pulse train as this sets the frequency comb linewidth to be less than the repetition frequency. This comb linewidth determines the ultimate frequency resolution for experiments using the VUV frequency comb~\cite{Marian}. The $\alpha$ parameters discussed above are a direct measure of the intensity-to-phase noise conversion in the below-threshold generation process that is directly relevant to the linewidth of a VUV frequency comb mode.  For instance, using $\alpha$ parameters extracted from the experiment, we can estimate the percentage of the power that would be removed from individual 1~kHz linewidth comb modes for a laser source with a 1\% RMS intensity fluctuations with a characteristic 1/f distribution for each of the generation processes~\cite{HallReview}. The results are presented in Table~1. For comparison, we include $\alpha$ and phase values for harmonics produced high above the ionization threshold, for instance in argon at an intensity of $2\times 10^{14}$~W/cm$^2$.

Our estimates of the intensity-to-phase noise put firm limits on the intensity noise that can be tolerated to preserve the comb structure in HHG, and it is clear from Table 1 that a very high degree of pulse-to-pulse coherence should be possible in principle.  To put this question to rest, we demonstrate pulse-to-pulse coherence of the 7th harmonic. For this study, the harmonic radiation is collimated and sent through a cross-correlation interferometer with a delay set so that a pulse in the harmonic pulse train will interfere with the subsequent pulse, allowing us to measure the degree of pulse-to-pulse coherence~\cite{Xu}. Since the 7th harmonic radiation is absorbed in air and most solid materials, extreme care was taken in the design and construction of this interferometer with all necessary optical alignments being conducted under vacuum.

Figure 3 shows the interference signal as the path length of the interferometer was scanned over 600 nm in about 2 seconds.  The pulse-to-pulse coherence is clearly visible in the figure as a cross correlation that corresponds to the 7th harmonic wavelength of $\approx$150 nm.  The contrast ratio shown in the figure is only $\approx 6\%$, however if we take into account imperfect collimation and severe power imbalance in the interferometer, we estimate that the maximum achievable contrast ratio was 10-15\%.  To provide a zero background for a significantly enhanced measurement contrast, we also used a lock-in detection by dithering the length of the interferometer at 392 kHz as the length of the interferometer was scanned. The demodulated signal is also shown in Fig. 3.

Since imperfect alignment in the interferometer and pulse overlap can easily decrease the contrast ratio further, our measurement provides only a lower bound on the coherence. By measuring the pulse-to-pulse coherence we place a lower bound on the coherence time ($\sim 10$ ns) which is roughly 5 orders of magnitude longer than that measured before~\cite{Cavalieri, Lynga}. The corresponding frequency comb linewidth is narrower than 20 MHz.

In summary, we have found experimental and theoretical evidence for multiple contributions to the generation of below-threshold harmonics. In particular, one of these contributions is dominated by laser-driven continuum dynamics, in analogy with the semi-classical model for generation of high order harmonics. Our measurements of the spatial and spectral profiles and of the intensity dependence of the interference pattern all confirm that the intensity-dependent phase is larger for these low-order harmonics than for high-order harmonics due to the influence of the atomic potential. We used our measured phase coefficients to estimate the coherence one can expect in both low-order and high-order harmonic frequency combs, and demonstrated that indeed, pulse-to-pulse coherence was maintained for the below-threshold harmonic generation.

\section*{Methods}
\noindent{\underline{The enhancement cavity}} \\
\noindent{To excite the enhancement cavity, we use a high-powered, frequency-controlled 1070~nm fiber laser system that supplies 100 fs pulses at 136 MHz with a pulse energy of 75 nJ~\cite{Hartl, Schibli}. When the pulse train is resonant with the cavity (obtained by actively controlling the two independent degrees of freedom of the frequency comb, the repetition frequency and the offset frequency), we achieve an enhancement of 260, which translates to an intracavity pulse energy of 19 $\mu$J.  Two 10-cm radius of curvature mirrors within the cavity produce a focal spot area of 960 $\mu$m$^2$ that yields a peak intensity of $4 \times 10^{13}$ W/cm$^2$. To generate harmonic radiation, we inject xenon gas near the intracavity focus using a glass nozzle with a 100 $\mu$m aperture and a backing pressure of 425 Torr.  An XUV diffraction grating is utilized as one element of the enhancement cavity so that a portion of the harmonic radiation generated diffracts out of the cavity and impinges on a fluorescent plate~\cite{grating}. By imaging the fluorescence onto a CCD camera, we are able to make measurements of the power and far field profile for individual harmonics. The exact location of the experimental laser focus relative to the gas jet, which is only known to within 200 $\mu$m, is inferred by aligning the experimental and theoretical result at the lowest intensity step for harmonic 7. This results in all the intensity step positions being remarkably consistent between theory and experiment.}

\noindent{\underline{Theoretical calculations}} \\
\noindent{In the semi-classical calculations, we performed a further check into the origin of the $\alpha_0 \approx 0$ phase contribution. We attempted to recover short trajectories with small phase coefficients and negative return energies by initiating electron trajectories with non-zero velocity at $x=0$, simulating their ionization as a multiphoton process. Under the limited range of initial conditions where we do find such trajectories, their phase coefficients are also larger than in the no-potential case with $\alpha_1 \approx 0.5\pi$. Fig. 2 shows that such  $\alpha_1$ contributions are not present in the fully quantum mechanical results. This confirms our conclusion that the $\alpha_0$ contribution is a multiphoton process.

To extract values for the phase coefficients $\alpha_0$ and $\alpha_2$ shown in Table 1, we made careful measurements of the inner and outer parts of the spatial (or spectral) profile of harmonic 7 at a well known intensity~\cite{He}.  If we assume that the harmonic has a Gaussian spatial (temporal) profile with an extra intensity dependent phase, and a focal spot (duration) which is one half of the IR spot (pulse), we find $\alpha_2 \approx 2.7 \pi$ and $\alpha_0 \leq 0.2 \pi$. Likewise, using the intensity separation between the two prominent steps in Fig. 1 as the ``period'' of the interference between the two contributions, we extract a value of $\alpha_2-\alpha_0 \approx 3 \pi$~\cite{Zair2008}.  All three of these measurements are thus in excellent agreement with the theoretical predictions.

\noindent{\underline{Coherence measurements}} \\
\noindent{For the pulse-to-pulse coherence measurements, the harmonic radiation outcoupled with the intracavity diffraction grating was sent through a VUV interferometer (see Fig. 4).  A MgF$_2$ beamsplitter, calcium fluoride lenses and aluminum mirrors optimized for operation in the VUV were used for the construction of the interferometer.  One end-mirror of the interferometer was mounted on a piezoelectric transducer (PZT) so that the optical delay of the interferometer could be scanned and quickly modulated.  To detect radiation at the output port, we used a conventional photomultiplier tube (PMT) with a thin layer of sodium salicylate on the front surface.  The sodium salicylate fluoresces at a wavelength of around 420 nm when struck by the 7th harmonic radiation, which matches very well to the peak detection efficiency of the PMT.  To observe the interference with a better contrast, we applied a 392 kHz modulation to the PZT.  The PZT and mounted mirror had a natural resonance close to this frequency allowing enhanced mirror travel. The demodulated signal is basically the first-order derivative of the DC-based interference measurement.}

$^*$Corresponding author: Ye@JILA.Colorado.edu
\begin{acknowledgments}
We gratefully thank I. Hartl, A. Marcinkevi\v{c}ius and M. Fermann at IMRA America, Inc. for the design and construction of the
high-power Yb-fiber laser system. Funding at JILA is provided by DARPA, NIST and NSF.
Funding at LSU is provided by the NSF through grants number PHY-0449235 and PHY-0701372, and by the CCT at LSU. KJS acknowledges support from the Ball Family Professorship. Portions of this research were conducted with high performance computational resources provided by the Louisiana Optical Network Initiative (http://www.loni.org).
\end{acknowledgments}

\newpage

\noindent{
\begin{table}
\caption{Frequency comb parameters for below threshold harmonics ($\hbar \omega < I_p $) at an intensity of $I=2\times10^{13}$ W/cm$^2$ and for above threshold harmonics ($\hbar \omega > I_p $) at $I=2\times10^{14}$ W/cm$^2$.  Also listed is the rms phase noise ($\phi_{rms}$) and an estimate of the power left in a 1 kHz carrier for an individual frequency comb mode with 1\% integrated intensity noise in frequencies above 1 kHz in a 1/f distribution.}
\begin{tabular}{llllll} \hline \hline
    & \multicolumn{2}{c}{$\hbar \omega < I_p$} & & \multicolumn{2}{c}{$\hbar \omega > I_p$} \\ \cline{2-3} \cline{5-6}
    Trajectory & $j=0$~~~ & $j=2$~~~ & & $j=1$~~~ & $j=2$~~~ \\
    $\alpha$  & $0.2\pi$ & $2.7\pi$ & & $0.2\pi$ & $2\pi$ \\
    $\phi_{rms} $ ~~~ & 0.01  & 0.16 & & 0.12 & 1.16 \\
    carrier power (\%) &  99.99 & 97.59 & & 98.67 & 26.08 \\ \hline \hline
  \end{tabular}

\end{table}}

\newpage

\begin{figure}[t]
\includegraphics[width=16cm]{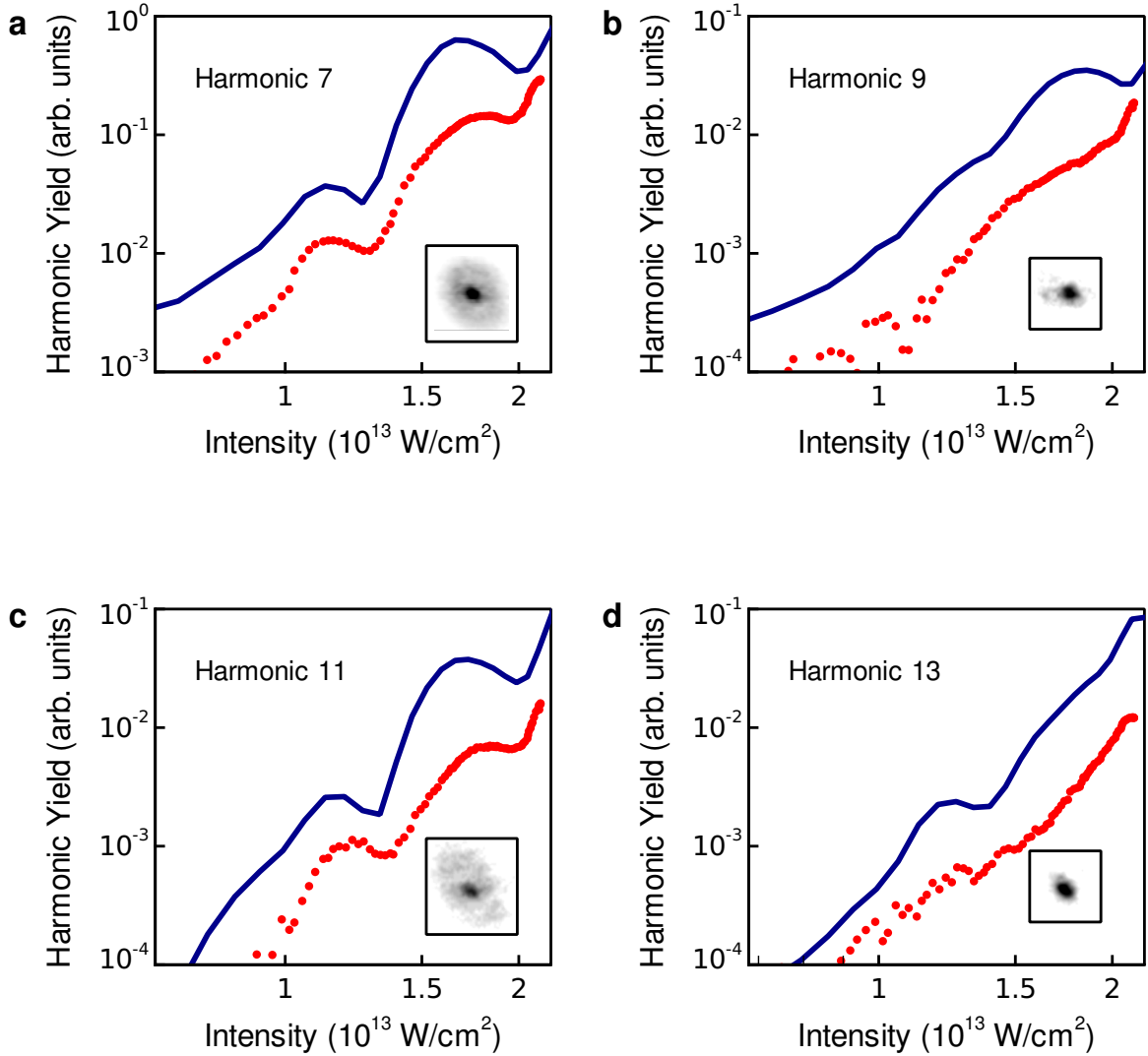}
\caption{Harmonic yield plotted as a function of the intensity at the center of the xenon jet.  (a)-(d) show harmonics 7 through 13, respectively.  The red filled circles are measured and the blue solid lines are theory (see text), both for the on-axis yield. Theoretical harmonic yields were scaled in magnitude to offer a clear comparison. The insets show the far-field spatial profile for each harmonic. Halos on the beam profile and strong oscillations in the intensity dependent yield are clearly visible in harmonics 7 and 11, providing clear evidence of multiple generation pathways with distinct intensity dependent phases.}

\label{theorydata}
\end{figure}

\begin{figure}[h]
\includegraphics[width=16cm]{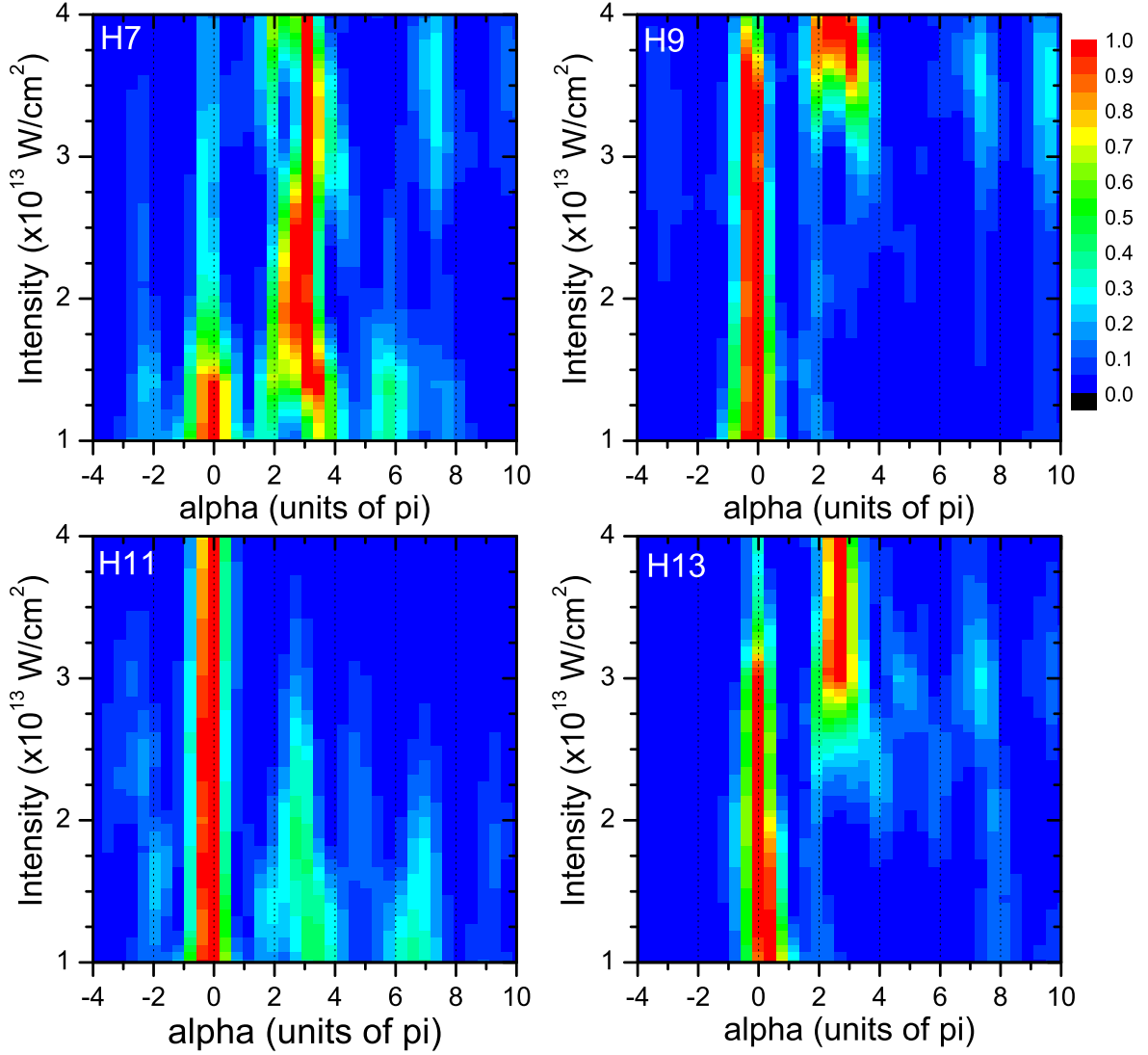}
\caption{Quantum path distributions calculated for harmonics 7-13. The color scale has been normalized for each intensity and shows only relative strengths.  Harmonics 7 and 11 show large quantum path contributions at $\alpha \approx 0$ and $\alpha \approx 3 \pi$ that manifest themselves in the spatial profiles with halos and intensity-dependent yields with large oscillations, as shown in Fig. 1.  Harmonics 9 and 13 only show prominent contributions at $\alpha \approx 0$ for intensities below $\sim 3 \times 10^{13}$ W/cm$^2$, so the features described for harmonics 7 and 11 are largely absent. }
\label{paths}
\end{figure}

\small
\begin{figure}[htb]
\centerline{
\includegraphics[width=16cm]{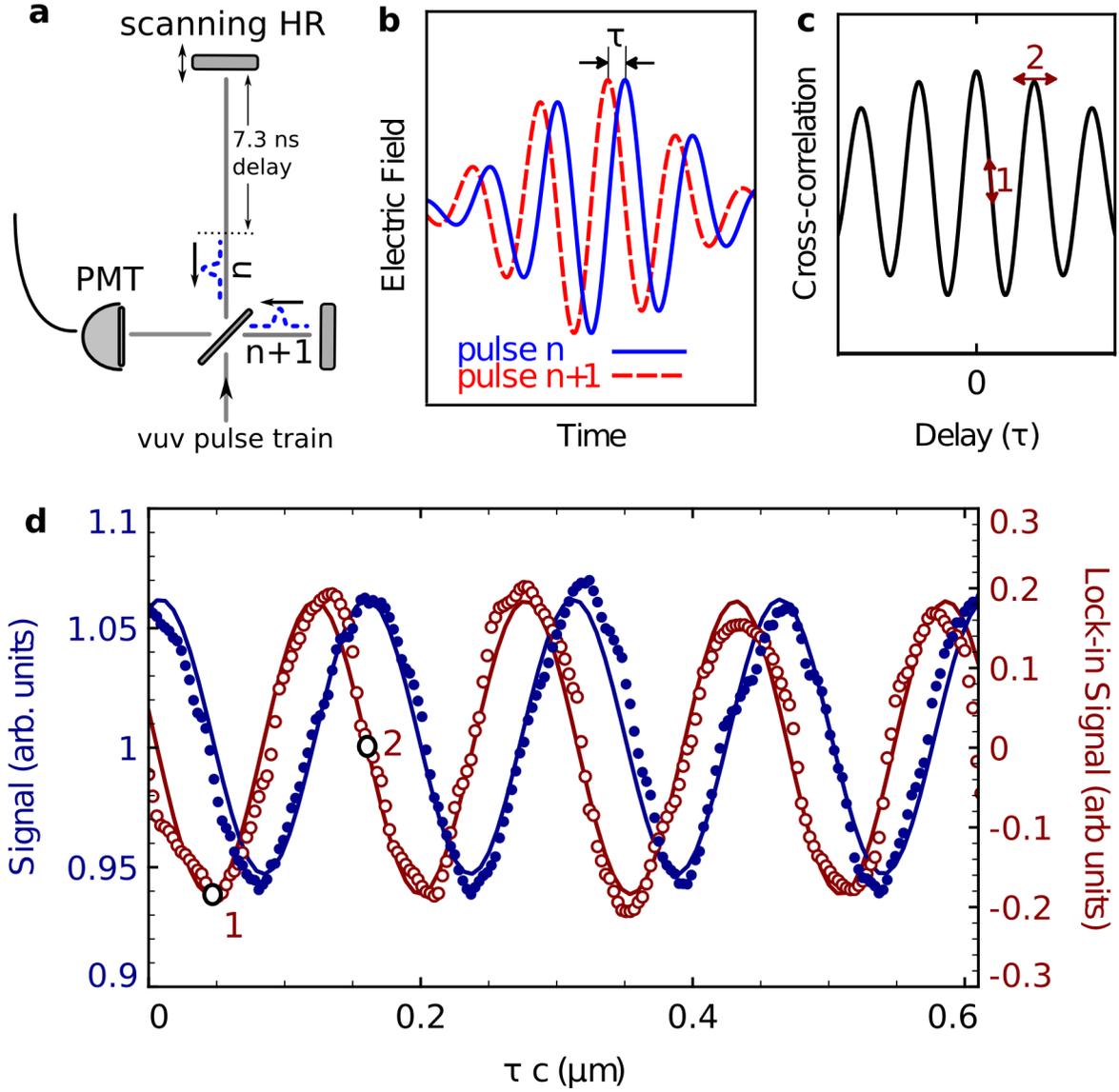}}
\caption{7th harmonic pulse-to-pulse coherence measurement. a) Pulse-to-pulse measurement setup: VUV interferometer with the delay set for pulse n in the 7th harmonic pulse train to interfere with pulse n+1. b) Figure depicting the electric field of pulse n in the 7th harmonic pulse train that has been sent through an optical delay line so that it interferes with the n+1 pulse in the train, which is shown as the dotted red trace. A cross-correlation signal results as the length of the delay line is scanned, which is shown in c).  d) The blue closed circles show our experimental measurement of the cross correlation as the path length of the interferometer is linearly scanned, showing a contrast ratio of $\approx 6\%$ limited by the spatial mode overlap and the power imbalance between the two interfering pulses. The red open circles display the signal achieved from a lock-in detection of a 392 kHz modulation of the delay line.  Both sets of data are fit with sinusoidal functions. Point 1 in c) and d) shows the positions of a maximum negative lock-in signal where the cross-correlation signal at DC has a maximum negative slope. Point 2 shows the position of a zero lock-in signal with the DC cross-correlation at its maximum.}
\label{beatnote}
\end{figure}
\normalsize

\small
\begin{figure}[htb]
\centerline{
\includegraphics[width=16cm]{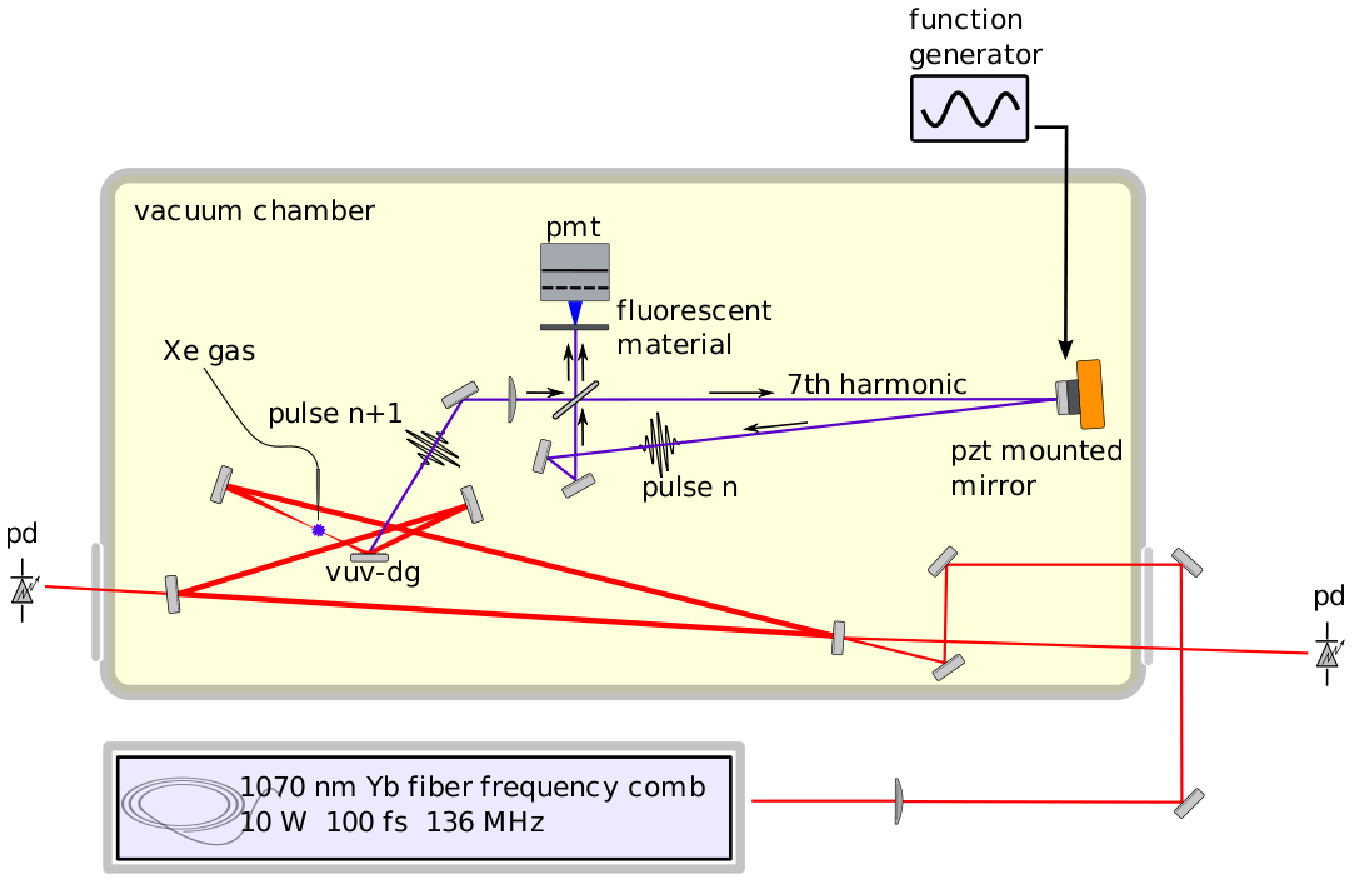}}
\caption{Experimental setup for the demonstration of VUV pulse-to-pulse coherence.  A cavity-enhanced Yb fiber frequency comb is used to generate harmonics in xenon gas. Photodiodes (pd) are used to monitor the reflected and transmitted fundamental light and maintain resonance between the frequency comb and the enhancement cavity~\cite{Jones1, Gohle}.  An intracavity VUV diffraction grating (vuv-dg) is used to outcouple the 7th harmonic radiation, which is then sent through an cross-correlation interferometer.  One mirror in the delay arm of the interferometer is mounted on a piezoelectric transducer (pzt), allowing scans of the differential path length.  The 7th harmonic radiation at the output of the interferometer is detected by monitoring the fluorescence from a sodium salicylate coated plate with a photomultiplier tube (pmt).}
\label{setup}
\end{figure}
\normalsize


\begin{thebibliography}{200}
\bibitem{footnote2} The calculations are performed with 50 fs pulses to save computing time. We have checked that the results do not change for longer pulses.
\end{thebibliography}
\end{document}